\documentclass[conference]{IEEEtran}

\usepackage{cite}
\usepackage{amssymb,amsfonts}
\usepackage{algorithmic}
\usepackage{graphicx}
\usepackage{textcomp}
\usepackage{xcolor}
\usepackage{todonotes}
\usepackage{amssymb}
\usepackage{booktabs}
\usepackage{amsmath}
\usepackage{listings}
\usepackage{romannum}
\usepackage{graphicx}
\usepackage{caption}
\usepackage{gnuplottex}

\usepackage{fancyhdr}
\usepackage[utf8]{inputenc}

\graphicspath{ {./images/} }

\usepackage{listings}
\usepackage{xcolor}

\definecolor{codegreen}{rgb}{0,0.6,0}
\definecolor{codegray}{rgb}{0,0,0}
\definecolor{codepurple}{rgb}{0.58,0,0.82}
\definecolor{backcolour}{rgb}{0.95,0.95,0.92}

\definecolor{blue-main}{rgb}{0,0,1}
\definecolor{dkgreen}{rgb}{0,0.6,0}
\definecolor{gray}{rgb}{0.5,0.5,0.5}
\definecolor{mauve}{rgb}{0.58,0,0.82}

\newcommand\blfootnote[1]{%
  \begingroup
  \renewcommand\thefootnote{}\footnote{#1}%
  \addtocounter{footnote}{-1}%
  \endgroup
}

\usepackage[pscoord]{eso-pic}
\newcommand{\placetextbox}[3]{
 \setbox0=\hbox{#3}
 \AddToShipoutPictureFG*{ \put(\LenToUnit{#1\paperwidth},\LenToUnit{#2\paperheight}){\vtop{{\null}\makebox[0pt][c]{#3}}}
 }
 }
 \placetextbox{.23}{.055}{\small{978-1-7281-6677-3/20/\$31.00~\copyright 2020 IEEE}}

\begin{document}

\title{
    Performance Evaluation of ParalleX Execution model on Arm-based Platforms\\
    \begin{large}
      A Preprint - \today
    \end{large}
}

\author{
\IEEEauthorblockN{Nikunj Gupta\IEEEauthorrefmark{1}\IEEEauthorrefmark{4}, Rohit
Ashiwal\IEEEauthorrefmark{1}, Bine
Brank\IEEEauthorrefmark{3}, Sateesh K. Peddoju\IEEEauthorrefmark{1}
Dirk Pleiter\IEEEauthorrefmark{3}}
\IEEEauthorblockA{\IEEEauthorrefmark{1}\textit{Dept. of CSE,} \textit{IIT Roorkee}, Roorkee, India, Email: gnikunj@cct.lsu.edu;\{rashiwal,sateesh\}@cs.iitr.ac.in}
\IEEEauthorblockA{\IEEEauthorrefmark{3}\textit{JSC}, \textit{Jülich Research Centre}, 52425 Jülich, Germany, Email: \{b.brank,d.pleiter\}@fz-juelich.de}
\IEEEauthorblockA{\IEEEauthorrefmark{4}The STE$||$AR Group, http://stellar-group.org}
}

\maketitle


\begin{abstract}
The HPC community shows a keen interest in creating diversity in the CPU ecosystem. The advent of Arm-based processors provides an alternative to the existing HPC ecosystem, which is primarily dominated by x86 processors. In this paper, we port an Asynchronous Many-Task runtime system based on the ParalleX model, i.e., High Performance ParalleX (HPX), and evaluate it on the Arm ecosystem with a suite of benchmarks. We wrote these benchmarks with an emphasis on vectorization and distributed scaling. We present the performance results on a variety of Arm processors and compare it with their x86 brethren from Intel. We show that the results obtained are equally good or better than their x86 brethren. Finally, we also discuss a few drawbacks of the present Arm ecosystem.
\end{abstract}

\begin{IEEEkeywords}
Asynchronous Many-Task, HPX, Parallel Computing, ParalleX
\end{IEEEkeywords}


\section{Introduction}

\blfootnote{\copyright 2020 IEEE.  Personal use of this material is permitted.  Permission from IEEE must be obtained for all other uses, in any current or future media, including reprinting/republishing this material for advertising or promotional purposes, creating new collective works, for resale or redistribution to servers or lists, or reuse of any copyrighted component of this work in other works.}The High-Performance ecosystem has shown a keen interest in shifting from the
traditional x86 architecture to Arm-based processors. The Japanese exascale
system called Fugaku~\cite{fugaku} is based on the A64FX processor from Fujitsu.
In Europe, the European Processor Initiative (EPI) is working on a processor
that will (among others) include Arm-based cores and is positioned as an HPC
technology. In the USA, Sandia National Laboratories has deployed a large-scale system based on
the ThunderX2 processor from Marvell, which is based on the Armv8 ISA.

This development triggers the question of whether the HPC software ecosystem is
ready for exploiting Arm. This concerns, in particular, a recent extension of
the Arm ISA, which is called Scalable Vector Extension
(SVE)~\cite{stephens2017arm}. SVE is vector-length agnostic, unlike AVX or
AVX512 from Intel, where SIMD width is fixed to 256 and 512 bit, respectively.
Size has some significant consequences when programming for SVE. For AVX and
AVX512 (and similar SIMD ISAs), data types that are defined (e.g., \_\_m512d
for a vector of eight doubles) have a size known at compile-time. For SVE, this
is a priori, not the case, as the vector length is only known at runtime.

These hardware complexities are matched with software complexities. Operating
systems, compiler, and library support are required to provide a functional
environment that supports large-scale HPC applications and ensure they can both
be easily ported to such new hardware and exploit it efficiently. One such class
of applications is the one exploiting parallelism through task-based
programming. Asynchronous Many-Task (AMT) runtime system models task-based
programming and offers an alternative to the conventional programming models like
Message Passing (MPI). In an AMT model, the program can be broken down into
tasks, with each task having a dependency on some other task generating a data
flow based process. During program execution, these tasks are launched
arbitrarily based on the input data and the DAG generated, enabling multiple
concurrent tasks running as computation kernels. The scheduler deals with the load
imbalance. These characteristics of the AMT model poses it as a viable
alternative in an era where future algorithms are expected to feature an increased
dynamic behavior and low uniformity. In this paper, we explore an AMT
model on Arm-based processors.

Section \Romannum{2} talks about Related Work in porting and evaluating the
Arm processors. Section \Romannum{3} discusses the ParalleX execution model and
HPX. Section \Romannum{4} touches on key concepts required to understand the
benchmarks and the results. Section \Romannum{5} describes the benchmarks and
system setup, and Section \Romannum{6} discusses the results.

\section{Related Work}

S. McIntosh-Smith et al.~\cite{McIntoshSmith2018ComparativeBO} recorded the
initial set of performance evaluations on mainstream Arm processors on large HPC
systems. These evaluations showcased performance and cost benefits for a class
of applications. The result considers performance on a single-node. They later released distributed application results in~\cite{mcintosh2019scaling}. Jackson et al.~\cite{10.1145/3324989.3325722}
investigated the performance of distributed memory communications (MPI), as well
as scientific applications utilizing MPI on ThunderX2. We also found that
Mont-Blanc project~\cite{banchelli2019mb3} investigated energy consumptions
during the execution of benchmarks and mini-apps.

While there has been decent research and performance evaluation on the
conventional computation models (such as OpenMP + MPI), there are no performance
numbers available for AMT runtime systems on mainstream Arm processors.

\subsection{Our contribution}

In this paper, we execute several benchmarks on an AMT based on the
ParalleX model, i.e. HPX. We investigate both distributed and shared memory
models with a special emphasis on vectorization. Furthermore, we provide
performance comparisons of mainstream Arm processors, such as Kunpeng 916
(HiSilicon), ThunderX2 (Marvell) and A64FX (Fujitsu), with their x86 brethren from
Intel.

\section{Background}

\subsection{ParalleX Execution Model}

ParalleX execution model~\cite{5364511} was devised to address the
critical bottlenecks of exascale HPC systems namely:

\begin{itemize}
    \item Starvation - insufficient parallelism
    \item Latencies - time-distance delay of remote resource accesses
    \item Overheads - extra work for management of parallel actions and resources on the critical path that are not necessary in the sequential variant
    \item Contention - delays due to lack of availability of resources
\end{itemize}

The ParalleX model offers an alternative to conventional computational models,
such as MPI.

ParalleX improves the efficiency of the application by reducing synchronization
and scheduling overheads. Resource utilization is achieved through increased
asynchrony. Contention overheads are significantly reduced by employing adaptive
scheduling and routing. This technique is instrumental in handling memory bank
conflicts. Data directed computing using message-driven computation and
lightweight synchronization mechanisms results in visible
scalability improvements, at least for certain classes of problems. ParalleX achieves power
reductions by reducing extraneous calculations and data movements.

\subsection{High Performance ParalleX Runtime System}

\begin{figure}[h]
    \centering
        \includegraphics[width=\linewidth]{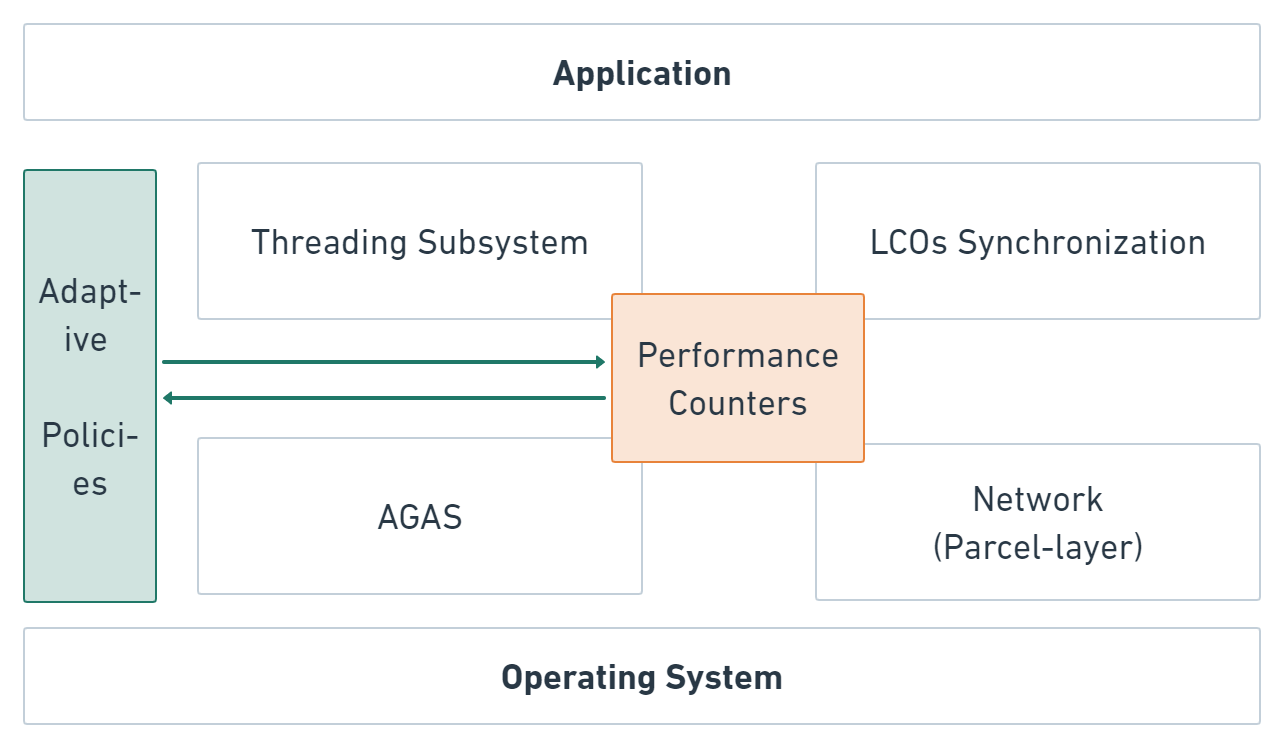}
        \caption{HPX Architecture Overview}
    \label{Fig:arch}
\end{figure}

High Performance ParalleX (HPX)~\cite{anderson2011application, Heller2013,
10.1145/2676870.2676883, Heller2017, hartmut_kaiser_2020_3675272} is the first
open-source implementation of ParalleX execution model. HPX exposes an ISO C++
standard confirming API, which enables wait-free asynchronous parallel programming,
including futures, channels, and other synchronization primitives. Being C++
standards conforming, HPX can run on a single machine as well as a cluster with
thousands of nodes. Figure \ref{Fig:arch} gives us the architecture of HPX.

HPX utilizes Active Global Address Space (AGAS) for addressing any HPX object
globally. Every object is assigned a Global Identifier (GID) that persists until
object destruction. Furthermore, AGAS supports load balancing through object
migration. HPX utilizes lightweight threads, called HPX threads, that are
executed and scheduled on top of OS threads. Local Control Object (LCO) is a
family of synchronization functions used to synchronize tasks generated by the
application. HPX has an active-message networking layer that ships functions to
the objects they operate on, i.e., the Parcel subsystem.

\subsection{Roofline Model}

The Roofline Model~\cite{10.1145/1498765.1498785, williams2008roofline, 10.5555/1713792}
provides performance estimation keeping in mind the bottlenecks and bounds to
predict a more realistic performance estimation. Usually, the performance is
expressed as a function of peak Computational Performance (CP) of the
architecture and the reachable peak I/O Bandwidth (BW). The CP is the maximum
number of floating-point operations that the processor can achieve.

The Arithmetic Intensity (AI) is the number of operations executed per byte
accessed from the main memory. It reveals the complexity of the algorithm. The model
identifies the CP or the I/O BW as limiting factors. Ergo, the maximal
attainable performance is always below the roofline obtained from both
parameters:

\begin{equation}
  Attainable\ Performance = min(CP,\ AI \times BW)
  \label{eq:roofline}
\end{equation}

\section{Stencil Codes}

In this paper, we explore a 1D stencil solver for the heat equation in one dimension
and a 2D stencil solver implementing a Jacobi solver.

\subsection{1D Heat Equation}

The diffusion equation, also known as heat equation in one dimension is given by:

\begin{equation}
    \frac{\partial u}{\partial t} = \alpha \frac{\partial^2 u}{\partial x^2}
\end{equation}

where $u(x,t)$ is unknown and $\alpha$ is the diffusion constant. On
discretizing the domain and replacing the derivatives by finite differences, one
can derive a 3-point stencil to update a single cell as:

\begin{equation}
    \begin{array}{l@{}l}
    T_{new}(x,y)
         &{} = T_{old}(x,y) \\
         &{} \\
         &{} + \alpha \frac{\displaystyle \Delta t}{\displaystyle \Delta x^2} [T_{old}(x-1,y) - 2 * T_{old}(x,y) \\
         &{} \\
         &{} + T_{old}(x+1,y)] \\
    \end{array}
    \label{eq:1d}
\end{equation}




\subsection{Jacobi Method for Linear Equation}


Jacobi solvers are a class of iterative solvers for linear equations given by:

\begin{center}
    $Ax\ =\ b,\ A \in \mathbb{R}^{NxN},\ x\ and\ b \in \mathbb{R}^N$
\end{center}

On a uniform grid with Dirichlet boundary conditions, the linear equation can be derived into a 5-point stencil to update a single cell as:


\begin{equation}
    \begin{array}{l@{}l}
    T_{new}(x,y)
         &{} = [T_{old}(x,y+1) + T_{old}(x,y-1) \\
         &{} + T_{old}(x+1,y) + T_{old}(x-1,y)]/4 \\
    \end{array}
\end{equation}

A detailed derivation is available at~\cite{doi:10.1137/1.9781611971538}.




\section{Benchmarks}

This Section discusses the implementation details of benchmarks, machine
architectures and HPX configuration.

\subsection{1D Stencil}

For the 1D stencil, we implement a fully distributed 1D heat equation solver. This
benchmark accurately measures the total execution time of the application. We report
application execution time over kernel performance to investigate how the
complete application scales on a distributed setting.

We run the application for two cases, i.e., weak and strong scaling. For weak
scaling, we start with 480 million stencil points, and another 480 million
stencil points are added for every node. For strong scaling, we benchmark with
1.2 billion stencil points. The benchmarks iterate over a hundred time steps.

\begin{lstlisting}[language=C++,caption= {1D stencil solver implemented in HPX}, label={1d_kernel}, tabsize=1]
  hpx::parallel::for_each(policy,begin(range),end(range),
    [&U, local_nx, nlp, t] (std::size_t i)
    {
      if (i == 0))
        stencil_update(U, 1, local_nx, t);
      else if (i == nlp-1))
        stencil_update(U, i * local_nx,
           (i + 1) * local_nx - 1, t);
      else if (i > 0 && i < nlp-1)
        stencil_update(U, i * local_nx,
           (i + 1) * local_nx, t);
    }
  );
\end{lstlisting}

Listing \ref{1d_kernel} shows our implementation of the 1D stencil kernel.
\lstinline{stencil_update} takes in the two grids and applies the operation defined in Equation~\ref{eq:1d}.

\subsection{2D stencil}
\label{subsection:2d_stencil}

For the 2D stencil, we implement a shared-memory based 2D stencil implementing
Jacobi method. We make use of explicit vectorization to compare performance
 with an auto vectorized scalar code from the compiler. Assuming
that the cache size is large enough to accommodate three
rows of the grid, three memory transfers have to be done for every iteration,
implying that for a double, a total of 24 Bytes are fetched from the main memory for every
Lattice Site Update (LUP). Similarly, for a float, a total of 12 Bytes are
fetched from the main memory. Thus, the Arithmetic Intensity (AI) for floats and doubles are
$1/12$ LUP/Byte and $1/24$ LUP/Byte, respectively. The low arithmetic intensity
makes the application memory bound for a broad class of processors.


\begin{lstlisting}[language=C++,caption= {Generic 2D stencil kernel implemented with HPX and NSIMD.  \texttt{Container} can be
  an STL vector of scalar types, or an STL vector of vector types.}, label={2d_kernel}, tabsize=1]
  template <typename Container>
  void stencil_update(array_t<Container>& U,size_t ny,size_t t)
  {
  Grid<Container>& curr = U[t % 2];
  Grid<Container>& next = U[(t + 1) % 2];

  size_t row_length = curr.row_size();

  #pragma unroll
  for (size_t nx = 1; nx < row_length-1; ++nx)
  {
    // Stencil operation
    next.in(nx, ny) = (curr.in(nx-1, ny) + curr.in(nx+1, ny) + curr.in(nx, ny-1) + curr.in(nx, ny+1)) * 0.25f;
  }

  // Maintain the halo in case of simd
  if (std::is_same<typename Container::value_type,nsimd::pack<typename get_type<typename Container::value_type>::type>>::value)
    helper<Container>::shuffle(next, ny);
  }

  // Call to stencil_update
  hpx::util::high_resolution_timer t;
  for (size_t t = 0; t < steps; ++t)
  {
    hpx::parallel::for_each(
      policy, begin(range), end(range),
      [&U, t] (size_t i)
      {
        stencil_update<Container>(U, i, t);
      });
  }
  t.elapsed();
\end{lstlisting}

Listing \ref{2d_kernel} shows parts of the C++ code written in HPX to generate a generic 2D stencil
kernel that supports all floating-point data types. \lstinline{Grid} is our custom class that
abstracts away the data layout of our stencil. We make use of the Virtual Node Scheme~\cite{boyle2015grid} to generate a SIMD data
layout of the stencil. To maintain this data layout, we
need to update the boundaries to keep consistent data. This is done by shuffling the halo vectors
to update according to the changes brought after executing the current time step (see Line 18). At Line 17, we use C++ type traits and our custom \lstinline{get_type} meta-class to identify if a type is scalar or vector.

\begin{table*}[tbh!]
  \centering
  \caption{Specification of the Arm and x86 nodes utilised in the benchmarks.}
  \label{tab:nodes}
  \begin{tabular}{l|lllll}
  \toprule
  \textbf{}                                   & Intel Xeon E5-2660 v3 & HiSilicon Kunpeng 916 & Marvell ThunderX2     & Fujitsu (FX1000) A64FX \\ \hline
  \midrule
  \textbf{Processor Clock Speed}              & 2.6GHz                & 2.4GHz                & 2.4GHz                & 2.2Ghz   \\ \hline
  \textbf{Cores per processors}               & 10                    & 64                    & 32                    & 48 (compute) + 4 (helper)   \\ \hline
  \textbf{Processors per node}              & 2                     & 1                     & 1                     & 1         \\ \hline
  \textbf{Threads per core}                   & 2                     & 1                     & 4                     & 1      \\ \hline
  \textbf{Vectorization}                      & Double AVX2 Pipeline  & Single NEON Pipeline  & Double NEON Pipeline  & Double SVE 512-bit     \\ \hline
  \textbf{Double Precision FLOPS per cycle}   & 16                    & 4                     & 8                     & 32    \\ \hline
  \textbf{Peak Performance in GFLOP/s}        & 832                   & 614                   & 1228                  & 3379     \\
  \bottomrule
  \end{tabular}
\end{table*}

We run the application with strong scaling since we are inclined to
investigate the performance of the kernel. We report performance numbers for a
grid size of $8192\times 131072$. The row size has been chosen such that it fits easily
in caches for our described assumptions to be true. Furthermore, the grid size
is chosen to be large enough in an attempt to keep all processing cores busy.
The benchmark iterates over a hundred time steps.

\section{System Setup}

We use the following clusters to access processors:

\renewcommand{\labelitemi}{\textendash}

\begin{itemize}
  \item Juawei prototype cluster, JSC. We use the cluster to access Intel Xeon
  E5 2660 v3 and HiSilicon Kunpeng 916 nodes.
    \item Sage prototype cluster, JSC. We use the cluster to access Marvell ThunderX2 nodes.
    \item A64FX prototype cluster, Fujitsu. We use the cluster to access Fujitsu FX1000 A64FX nodes.
  \end{itemize}

Table \ref{tab:nodes} lists the specification for all processors. Table
\ref{tab:conFig} gives an overview of the dependencies of
our benchmark. We choose GCC over Arm HPC compiler and Fujitsu Compiler. Arm
compilers implement SVE data types using \lstinline{__sizeless_struct}.
\lstinline{__sizeless_struct} determines vector length at runtime, thus making
the SVE code portable in the sense that an executable can run all vector lengths
supported by SVE (i.e., 128bits through 2048bits). However, one cannot wrap
\lstinline{__sizeless_struct} into a struct or a class unless the struct is also
defined as \lstinline{__sizeless_struct}. Since we make heavy use of the C++ STL
vector and our custom class, we have to determine type and length at
compile time. As SVE is bleeding-edge technology, there is no synergy between
the implementation details of various compilers. Currently, only GCC provides
the choice of passing SVE vector length at compile time. Therefore, we use GCC.

\begin{table}[tbh!]
  \centering
  \caption{Benchmark dependencies Configuration}
  \label{tab:conFig}
  \begin{tabular}{l|ll}
    \toprule
    \textbf{Package Name}       & \textbf{Version}  \\ \hline
    \midrule
    GCC                         & 10.1              \\ \hline
    hwloc                       & 2.1               \\ \hline
    jemalloc                    & 5.2.1               \\ \hline
    boost                       & 1.66              \\ \hline
    HPX                         & commit c62d992    \\ \hline
    NSIMD                       & commit d4f9fc5    \\ \hline
    PAPI                       & 6.0.0    \\ \hline
    \bottomrule
  \end{tabular}
\end{table}

We make use of NSIMD\footnote{https://github.com/agenium-scale/nsimd} for explicit vectorization. We chose NSIMD over explicitly
using vector intrinsics because NSIMD provides a portable code supporting a
wide array of vectorizations, including SVE. Portability allowed us to use the
same code on all machine architectures and underlying vectors. To our knowledge,
NSIMD and Inastemp~\cite{Bramas2017} are the only C++ libraries for explicit vectorization that
support SVE data types.

\begin{figure}[h!]
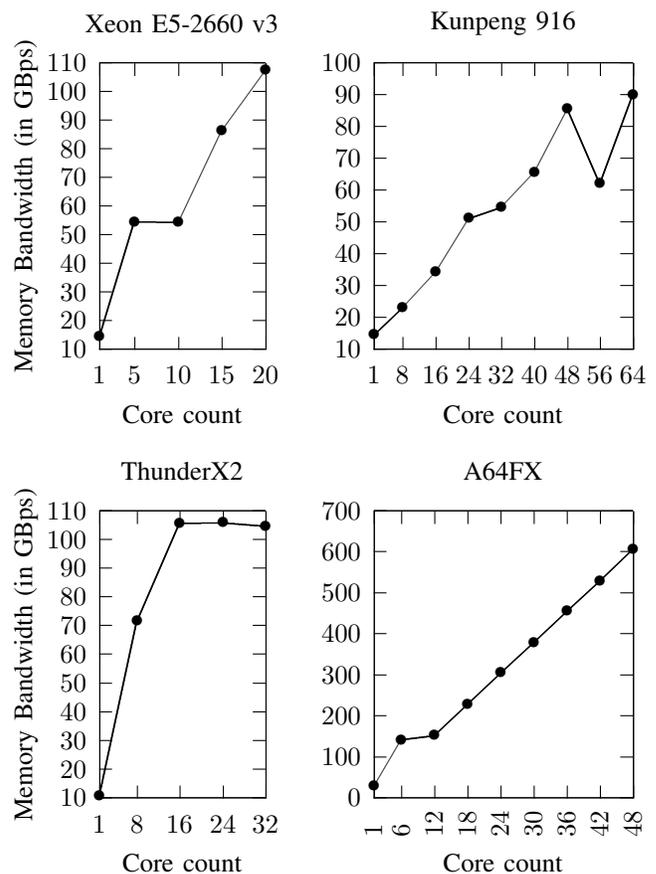

    \centering
    \captionsetup{justification=centering}
    \begin{gnuplot}[terminal=latex]
        set terminal latex rotate size 3.5in,5in

        set ylabel "Memory Bandwidth (in GBps)"
        set xlabel "Core count"

        set size 1,1
        set multiplot layout 2,2 title "Memory Bandwidth using STREAM COPY Benchmark"

        set title "Xeon E5-2660 v3"
        unset key

        set size 0.45,0.47
        set origin 0,0.47
        set xtics (1, 5, 10, 15, 20)

        plot \
        "plots/2d_stencil/x86/x86_64_bandwidth.dat" using 1:($2/1000) lw 2 lt 7 lc -1 w lp, \

        set title "Kunpeng 916"
        unset ylabel

        set size 0.55,0.47
        set origin 0.45,0.47
        set xtics (1, 8, 16, 24, 32, 40, 48, 56, 64)

        plot \
        "plots/2d_stencil/hi1616/aarch64_bandwidth.dat" using 1:($2/1000) lw 2 lt 7 lc -1 w lp, \

        set ylabel "Memory Bandwidth (in GBps)"

        set title "ThunderX2"

        set size 0.45,0.47
        set origin 0,0
        set xtics (1, 8, 16, 24, 32)

        plot \
        "plots/2d_stencil/x2/aarch64_bandwidth.dat" using 1:($2/1000) lw 2 lt 7 lc -1 w lp, \

        unset ylabel
        set title "A64FX"
        unset ylabel

        set size 0.55,0.47
        set origin 0.45,0
        set xtics (1, 6, 12, 18, 24, 30, 36, 42, 48) rotate by 90 offset 0,-1

        plot \
        "plots/2d_stencil/a64fx-new/131072/aarch64_bandwidth.dat" using 1:($2/1000) lw 2 lt 7 lc -1 w lp, \

        unset multiplot
    \end{gnuplot}
    \caption{Memory Bandwidth results using the STREAM COPY Benchmark with an array size of 128 million elements}
  \label{graph:stream}
\end{figure}

Apart from the specifications in Table \ref{tab:nodes}, we measure the memory
bandwidth using the STREAM COPY benchmark~\cite{McCalpin1995}. Results of the
run are available in Figure~\ref{graph:stream}\footnote{Higher STREAM benchmark results can be obtained on A64FX using Fujitsu compiler and special cache initialization techniques. We use the STREAM benchmark as a reference performance metric for our stencil application. For us to make assertions about the optimality of stencil codes, we need to make sure that we do not optimize STREAM benchmarks using techniques that cannot be applied to the 2D stencil code. We make sure that STREAM benchmarks are NUMA aware as we make our 2D stencil code NUMA aware.}.
The benchmark was run ten times, and the highest memory bandwidth for the core count is reported.
Due to the memory bound nature of stencil codes,
memory bandwidth results are important to determine how our results compare to
the expected peak deriving from the STREAM benchmark.

We run each variant of the 1D stencil and 2D stencil for three and five times respectively.
In case of 1D stencil, we report the least time consumed amongst all runs. For
2D stencil, we report the maximum performance achieved for a particular data type.
While hyperthreading is enabled on all the cores, we pin to the physical
PUs to ensure that the benchmark effectively uses L1 and L2 caches. In a
hyperthreaded scenario, the pressure on the cache increases that may result in cache evictions leading to a possible loss in performance. All the benchmarks are run
by pinning one thread per core using \lstinline{hwloc-bind}.
Finally, we turn all optimization flags on, i.e.
\lstinline{-O3, -ftree-vectorize, -ffast-math} for the best possible performances.

\textbf{Hardware Counters:} We use Linux \lstinline{perf} and PAPI to get access to
the hardware counters to better explain any aberration in results. All hardware
counters were run on a single physical core on a smaller grid size of 8192 $\times$
16384 for a hundred iteration.

\section{Results}
\label{sec:results}

\subsection{1D Stencil}

\begin{figure}[tbph!]
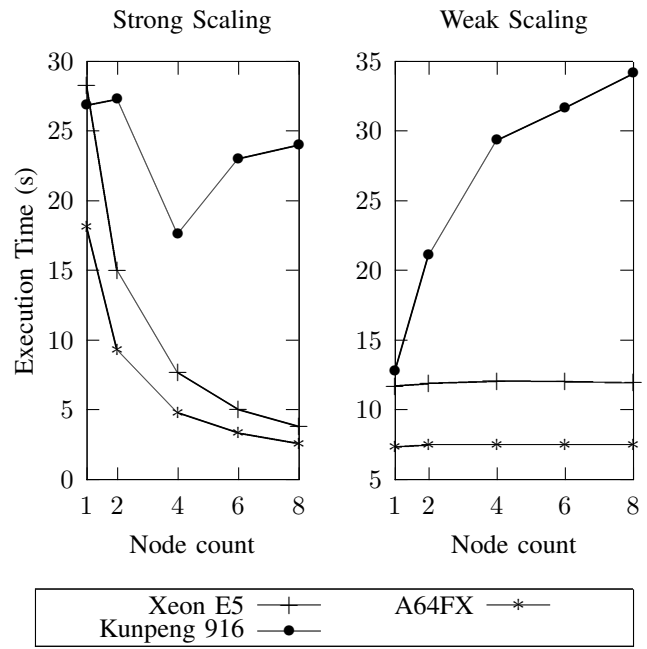

    \centering
    \captionsetup{justification=centering}
    \begin{gnuplot}[terminal=latex]
        set terminal latex rotate size 3.5in,3.8in

        set ylabel "Execution Time (s)"
        set xlabel "Node count"

        set size 1,1
        set multiplot layout 2,2 title "1D Stencil: Distributed Results"

        unset key

        set size 0.5,0.8
        set origin 0,0.1

        set title "Strong Scaling"

        set xtics (1, 2, 4, 6, 8)

        plot \
        "plots/1d_stencil/x86/result_x86" title "Xeon E5" lw 2 lc -1 w lp, \
        "plots/1d_stencil/hi1616/result_hi1616" title "Kunpeng 916" lw 2 lt 7 lc -1 w lp, \
        "plots/1d_stencil/a64fx/strong_a64fx" title "A64FX" lw 2 lt 9 lc -1 w lp, \

        set size 0.5,0.8
        set origin 0.5,0.1

        set title "Weak Scaling"

        unset ylabel

        set xtics (1, 2, 4, 6, 8)

        plot \
        "plots/1d_stencil/x86/weak_x86" title "Xeon E5" lw 2 lc -1 w lp, \
        "plots/1d_stencil/hi1616/weak_hi1616" title "Kunpeng 916" lw 2 lt 7 lc -1 w lp, \
        "plots/1d_stencil/a64fx/weak_a64fx" title "A64FX" lw 2 lt 9 lc -1 w lp, \

        set size 1,0.15
        set origin 0,-0.05
        unset title
        set key horizontal center
        unset tics
        unset xlabel
        unset ylabel
        set yrange [0:1]
        plot 2 t 'Xeon E5' lw 2 lc -1 w lp, \
             2 t 'Kunpeng 916' lw 2 lt 7 lc -1 w lp, \
             2 t 'A64FX' lw 2 lt 9 lc -1 w lp, \

        unset multiplot
    \end{gnuplot}
    \caption{1D stencil: strong and weak scaling results. Strong scaling is done over 1.2 billion stencil points. Weak scaling is done by adding 480 million stencil points per node.}
  \label{graph:strong}
\end{figure}

Figure~\ref{graph:strong} succinctly presents the results of strong and weak
scaling. The distributed application is implemented such that network latencies
can be hidden under compute. Furthermore, the application is also NUMA aware.
This is made possible by utilizing block allocators implemented within HPX. The
allocator allocates memory based on Linux's first touch data placement policy.
This is similar to OpenMP's \lstinline{schedule(static)} policy. Combined with
the block executor, we make sure that an HPX thread always spawns at a location
of data. Thus, we are able to make up for the lack of bandwidth between
chip-to-chip communications. For Fujitsu A64FX, we compiled all dependencies
using the Fujitsu compiler. Furthermore, we build the HPX parcelport backend
using Fujitsu MPI for easier integration with the system. For all other
processors, results are considered according to the benchmark dependencies, as
described in Table~\ref{tab:conFig}.

For Intel Xeon E5 and Fujitsu A64FX under strong scaling, the application takes
28s and 18s respectively for a single node and 3.8s and 2.5s respectively
involving eight nodes, which is close to linear scaling (the factor being 7.36 and
7.2, respectively). We expected a similar behavior as sufficient
parallelism can be derived from the given number of stencil points to hide
network latencies. Under weak scaling, the application takes 12s and 7.5s
respectively irrespective of the number of nodes which proves that the network
latencies are aptly hidden.

For Kunpeng 916, we do not observe linear scaling. On closer inspection, we
observed that the network performance on the Hi1616 nodes is unsatisfactory and that
the processor is not able to exploit the capabilities of the InfiniBand
network making it difficult to hide network latencies and the results are
transferred to the graph. This hypothesis is well proven under weak scaling,
where we see a significant increase in execution times as we increase the number
of nodes.

\subsection{2D Stencil}


The expected peak performance can be calculated with the recorded Memory
Bandwidth, and calculated Arithmetic Intensity
(See~\ref{subsection:2d_stencil}) by putting the values in Equation~\ref{eq:roofline}. We do not aim to be optimal as further
micro-optimizations can be made on our 2D stencil implementation.
Our results will be considered optimal if we reach close to this expected peak performance. We make use of hardware performance counters to explain any aberrations in the results.

\begin{figure}[tb]
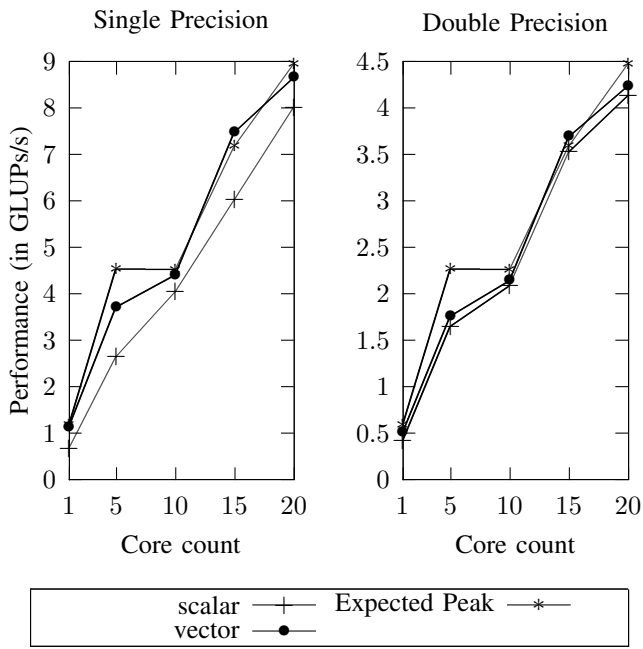

    \centering
    \captionsetup{justification=centering}
    \begin{gnuplot}[terminal=latex]
        set terminal latex rotate size 3.5in,3.8in

        set ylabel "Performance (in GLUPs/s)"
        set xlabel "Core count"

        set size 1,1
        set multiplot layout 2,2 title "2D Stencil: Intel Xeon E5-2660 v3"

        unset key

        set size 0.5,0.8
        set origin 0,0.1

        set title "Single Precision"

        set xtics (1, 5, 10, 15, 20)

        plot \
        "plots/2d_stencil/x86/x86_64_float" using 1:($3/1000) title "scalar" lw 2 lc -1 w lp, \
        "plots/2d_stencil/x86/x86_64_nfloat" using 1:($3/1000) title "vector" lw 2 lt 7 lc -1 w lp, \
        "plots/2d_stencil/x86/x86_64_bandwidth.dat" using 1:($2/12000) title "Expected peak" lw 2 lt 9 lc -1 w lp, \

        set size 0.5,0.8
        set origin 0.5,0.1

        set title "Double Precision"

        unset ylabel

        set xtics (1, 5, 10, 15, 20)

        plot \
        "plots/2d_stencil/x86/x86_64_double" using 1:($3/1000) title "scalar" lw 2 lc -1 w lp, \
        "plots/2d_stencil/x86/x86_64_ndouble" using 1:($3/1000) title "vector" lw 2 lt 7 lc -1 w lp, \
        "plots/2d_stencil/x86/x86_64_bandwidth.dat" using 1:($2/24000) title "Expected peak" lw 2 lt 9 lc -1 w lp, \

        set size 1,0.15
        set origin 0,-0.05
        unset title
        set key horizontal center
        unset tics
        unset xlabel
        unset ylabel
        set yrange [0:1]
        plot 2 t 'scalar' lw 2 lc -1 w lp, \
             2 t 'vector' lw 2 lt 7 lc -1 w lp, \
             2 t 'Expected Peak' lw 2 lt 9 lc -1 w lp, \

        unset multiplot
    \end{gnuplot}
    \caption{2D stencil: Results for Intel Xeon E5-2660 v3 with a grid size of 8192$\times$131072 iterated over 100 time steps}
    \label{graph:x86}
\end{figure}

Our noticeable observation about GCC is its ability to auto vectorize our 2D stencil
application. We observe that the instruction count with auto vectorized codes
is similar to our explicitly vectorized codes, sometimes even beating it at
instruction count. Visible differences come from cache-misses and CPU stalls arising from
different data layout pattern design by GCC and our code. The next few
paragraphs discuss processor specific results in detail.

Table~\ref{tab:perf-x86} describes the major contributing hardware counters that directly
contribute to performance differences for Intel Xeon E5. We observed a 2x difference in
instruction count between scalar and vector types, i.e., GCC is not able to
auto vectorize the code very well. The lower instruction count certainly helps an
explicitly vectorized code to perform better. Interestingly, the cache friendly
optimization from GCC is highly optimized for x86 architecture. This can be seen by
the lower cache miss counts for auto vectorized codes.

\begin{table}[h!]
  \centering
  \caption{Hardware Counters for Intel Xeon E5-2660v3}
  \label{tab:perf-x86}
  \begin{tabular}{l|ll}
    \toprule
    \textbf{Data Type}       & \textbf{Instruction} & \textbf{Cache Misses}  \\ \hline
    \midrule
    Float                   & 3.153$\times 10^{10}$     &  2.121$\times 10^{8}$       \\ \hline
    Vector Float            & 1.783$\times 10^{10}$    &  3.706$\times 10^{8}$        \\ \hline
    Double                  & 6.01$\times 10^{10}$    &  4.74$\times 10^{8}$       \\ \hline
    Vector Double           & 3.507$\times 10^{10}$    &  8.751$\times 10^{8}$       \\ \hline
    \bottomrule
  \end{tabular}
\end{table}

Furthermore, PUs are able to keep only certain memory transactions in the flight.
One can make use of CPU stall cycles to determine these numbers. Unfortunately, Intel
Xeon E5 2660v3 doesn't support these counters. Nonetheless, we believe that the lower instruction counts
relieve the memory controllers of excessive memory transactions improving the
performance further. We hypothesize that the improvements of up to 50\% with vectorized floats are a result of decreased memory transactions.
Given doubles occupy eight bytes, we do not expect much performance improvements with decreased instruction count on an already busy memory bus, which can be easily visualized
in the graphs with only up to 10\% improvements in performances.


\begin{figure}[tb]
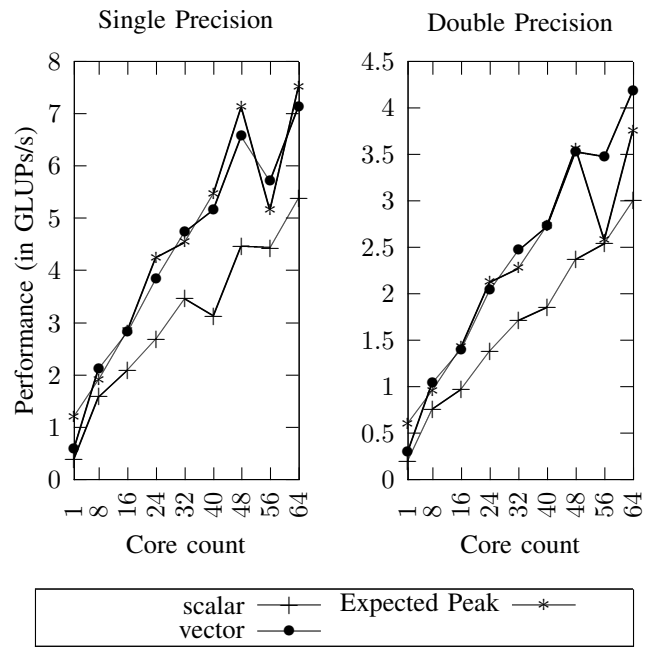

    \centering
    \captionsetup{justification=centering}
    \begin{gnuplot}[terminal=latex]
        set terminal latex rotate size 3.5in,3.8in

        set ylabel "Performance (in GLUPs/s)"
        set xlabel "Core count"

        set size 1,1
        set multiplot layout 2,2 title "2D Stencil: Huawei Hi1616"

        unset key

        set size 0.5,0.8
        set origin 0,0.1

        set title "Single Precision"

        set xtics (1, 8, 16, 24, 32, 40, 48, 56, 64) rotate by 90 offset 0,-1

        plot \
        "plots/2d_stencil/hi1616/Hi1616_float" using 1:($3/1000) title "scalar" lw 2 lc -1 w lp, \
        "plots/2d_stencil/hi1616/Hi1616_nfloat" using 1:($3/1000) title "vector" lw 2 lt 7 lc -1 w lp, \
        "plots/2d_stencil/hi1616/aarch64_bandwidth.dat" using 1:($2/12000) title "Expected peak" lw 2 lt 9 lc -1 w lp, \

        set size 0.5,0.8
        set origin 0.5,0.1

        set title "Double Precision"

        unset ylabel

        set xtics (1, 8, 16, 24, 32, 40, 48, 56, 64) rotate by 90 offset 0,-1

        plot \
        "plots/2d_stencil/hi1616/Hi1616_double" using 1:($3/1000) title "scalar" lw 2 lc -1 w lp, \
        "plots/2d_stencil/hi1616/Hi1616_ndouble" using 1:($3/1000) title "vector" lw 2 lt 7 lc -1 w lp, \
        "plots/2d_stencil/hi1616/aarch64_bandwidth.dat" using 1:($2/24000) title "Expected peak" lw 2 lt 9 lc -1 w lp, \

        set size 1,0.15
        set origin 0,-0.05
        unset title
        set key horizontal center
        unset tics
        unset xlabel
        unset ylabel
        set yrange [0:1]
        plot 2 t 'scalar' lw 2 lc -1 w lp, \
             2 t 'vector' lw 2 lt 7 lc -1 w lp, \
             2 t 'Expected Peak' lw 2 lt 9 lc -1 w lp, \

        unset multiplot
    \end{gnuplot}
    \caption{2D stencil: Results for Huawei Kunpeng 916 with a grid size of 8192$\times$131072 iterated over 100 time steps}
    \label{graph:hi1616}
\end{figure}

\begin{table}[tbh!]
  \centering
  \caption{Hardware Counters for HiSilicon Hi1616}
  \label{tab:perf-hi1616}
  \begin{tabular}{l|ll}
    \toprule
    \textbf{Data Type}       & \textbf{Instruction} & \textbf{Cache Misses}  \\ \hline
    \midrule
    Float                   & 4.3$\times 10^{10}$    &  3.148$\times 10^{9}$       \\ \hline
    Vector Float            & 4.144$\times 10^{10}$    &  2.512$\times 10^{9}$        \\ \hline
    Double                  & 8.321$\times 10^{10}$    &  5.639$\times 10^{9}$       \\ \hline
    Vector Double           & 8.236$\times 10^{10}$    &  4.953$\times 10^{9}$       \\ \hline
    \bottomrule
  \end{tabular}
\end{table}

\begin{figure}[tb]
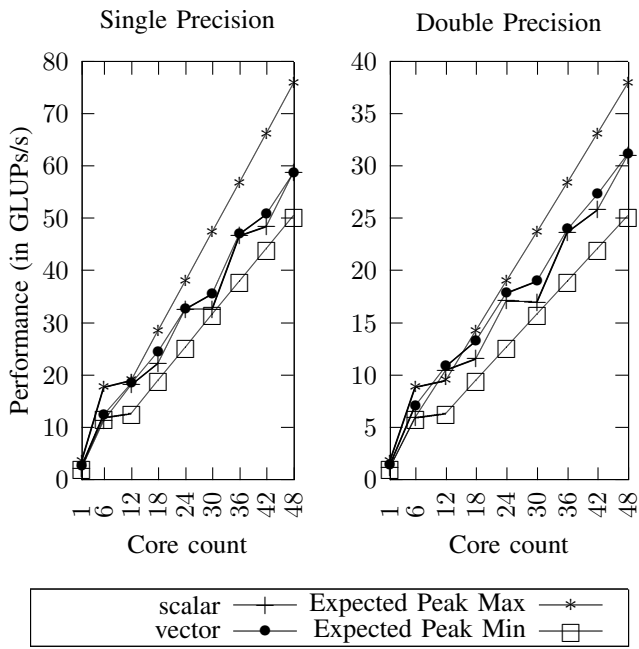

    \centering
    \captionsetup{justification=centering}
    \begin{gnuplot}[terminal=latex]
        set terminal latex rotate size 3.5in,3.8in

        set ylabel "Performance (in GLUPs/s)"
        set xlabel "Core count"

        set size 1,1
        set multiplot layout 2,2 title "2D Stencil: Fujitsu A64FX (Compute cores only)"

        unset key

        set size 0.5,0.8
        set origin 0,0.1

        set title "Single Precision"

        set xtics (1, 6, 12, 18, 24, 30, 36, 42, 48) rotate by 90 offset 0,-1

        plot \
        "plots/2d_stencil/a64fx-new/131072/a64fx_float" using 1:($3/1000) title "scalar" lw 2 lc -1 w lp, \
        "plots/2d_stencil/a64fx-new/131072/a64fx_nfloat" using 1:($3/1000) title "vector" lw 2 lt 7 lc -1 w lp, \
        "plots/2d_stencil/a64fx-new/131072/aarch64_bandwidth.dat" using 1:($2/8000) title "Expected peak Max" lw 2 lt 9 lc -1 w lp, \
        "plots/2d_stencil/a64fx-new/131072/aarch64_bandwidth.dat" using 1:($2/12000) title "Expected peak Min" lw 2 lt 9 lc -1 w lp, \

        set size 0.5,0.8
        set origin 0.5,0.1

        set title "Double Precision"

        unset ylabel

        set xtics (1, 6, 12, 18, 24, 30, 36, 42, 48) rotate by 90 offset 0,-1

        plot \
        "plots/2d_stencil/a64fx-new/131072/a64fx_double" using 1:($3/1000) title "scalar" lw 2 lc -1 w lp, \
        "plots/2d_stencil/a64fx-new/131072/a64fx_ndouble" using 1:($3/1000) title "vector" lw 2 lt 7 lc -1 w lp, \
        "plots/2d_stencil/a64fx-new/131072/aarch64_bandwidth.dat" using 1:($2/16000) title "Expected peak Max" lw 2 lt 9 lc -1 w lp, \
        "plots/2d_stencil/a64fx-new/131072/aarch64_bandwidth.dat" using 1:($2/24000) title "Expected peak Min" lw 2 lt 9 lc -1 w lp, \

        set size 1,0.15
        set origin 0,-0.05
        unset title
        set key horizontal center right
        unset tics
        unset xlabel
        unset ylabel
        set yrange [0:1]
        plot 2 t 'scalar' lw 2 lc -1 w lp, \
             2 t 'vector' lw 2 lt 7 lc -1 w lp, \
             2 t 'Expected Peak Max' lw 2 lt 9 lc -1 w lp, \
             2 t 'Expected Peak Min' lw 2 lt 9 lc -1 w lp, \

        unset multiplot
    \end{gnuplot}
    \caption{2D stencil: Results for Fujitsu A64FX with a grid size of 8192$\times$131072 iterated over 100 time steps. Expected Peak Max assumes two memory transfers per iteration and Expected Peak Min assumes three memory transfers per iteration.}
    \label{graph:a64fx}
\end{figure}

HiSilicon Hi1616 shows up to 80\% improvements with explicit vectorization.
Table~\ref{tab:perf-hi1616} describes differentiating hardware counter results.
The table suggests that GCC is able to
auto vectorize the results very well. Explicit vectorization resulted in a mere
5\% improvement in instruction count. The auto vectorization does fail in
exploiting caches that can be observed by a 10-20\% decline in cache misses by
moving to an explicitly vectorized code. We believe that the remaining difference
arises from Backend stalls. Unfortunately, Hi1616 doesn't support
CPU stall counters, but we hypothesize this based on our results on ThunderX2
(result described in the coming paragraphs) where we see similar behaviors.
Backend stalls are generally caused either by many
long-latency operations such as multiply and divide, or by long-latency memory operations, or both.
In our case, our stencil kernel uses simpler operations, so, a majority of such stalls are
caused by memory operations.
With explicit vectorization, we believe that we've reduced these memory transactions
considerably allowing us for better performances.


Another interesting observation is the sudden decrease in performance while
moving from 32 to 40 cores and 56 to 64 cores. It can be explained by
understanding how memory controllers work. When the application runs on 40
cores, two of the NUMA domains are fully saturated with respect to memory
bandwidth, while the third NUMA domain is only partially saturated. This uneven
distribution leads to faster iteration in the fully saturated domains, leaving a
trail of poor performance for the partially saturated domain. Therefore, a
partially saturated NUMA domain becomes the critical path for the benchmark.
With lower memory bandwidth available to the partially saturated domain, we see
a loss in performance.

For Fujitsu A64FX, the execution time for the 2D stencil kernel is less than 2s
for scalar and vector floats and about 3.5s for scalar and vector doubles while
utilizing all 48 compute cores.
Compared to the other processors, the execution time is significantly lower.
This piqued our interest, and we investigated whether we have enough parallelism
for HPX to take advantage of. Like every AMT model, HPX is known to have
contention overheads when the grain size is too small.


\begin{figure}[tb]
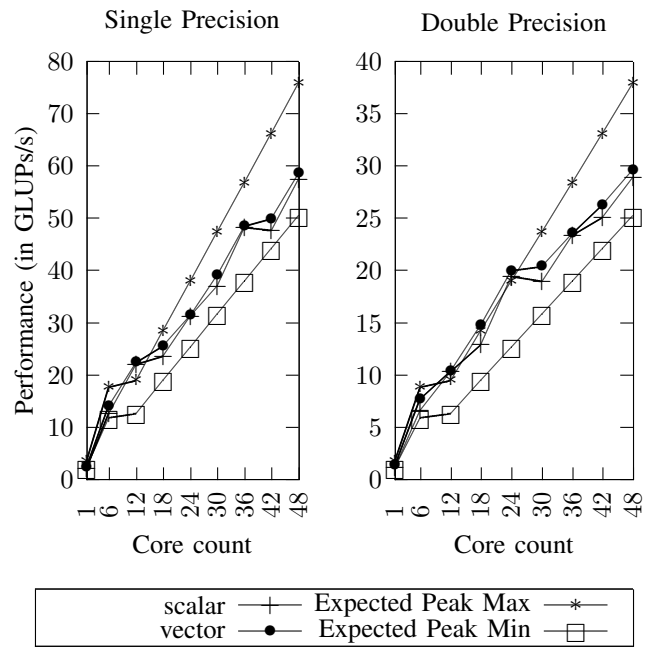

    \centering
    \captionsetup{justification=centering}
    \begin{gnuplot}[terminal=latex]
        set terminal latex rotate size 3.5in,3.8in

        set ylabel "Performance (in GLUPs/s)"
        set xlabel "Core count"

        set size 1,1
        set multiplot layout 2,2 title "2D Stencil: Fujitsu A64FX (Compute cores only)"

        unset key

        set size 0.5,0.8
        set origin 0,0.1

        set title "Single Precision"

        set xtics (1, 6, 12, 18, 24, 30, 36, 42, 48) rotate by 90 offset 0,-1

        plot \
        "plots/2d_stencil/a64fx-new/196608/a64fx_float" using 1:($3/1000) title "scalar" lw 2 lc -1 w lp, \
        "plots/2d_stencil/a64fx-new/196608/a64fx_nfloat" using 1:($3/1000) title "vector" lw 2 lt 7 lc -1 w lp, \
        "plots/2d_stencil/a64fx-new/196608/aarch64_bandwidth.dat" using 1:($2/8000) title "Expected peak Max" lw 2 lt 9 lc -1 w lp, \
        "plots/2d_stencil/a64fx-new/196608/aarch64_bandwidth.dat" using 1:($2/12000) title "Expected peak Min" lw 2 lt 9 lc -1 w lp, \

        set size 0.5,0.8
        set origin 0.5,0.1

        set title "Double Precision"

        unset ylabel

        set xtics (1, 6, 12, 18, 24, 30, 36, 42, 48) rotate by 90 offset 0,-1

        plot \
        "plots/2d_stencil/a64fx-new/196608/a64fx_double" using 1:($3/1000) title "scalar" lw 2 lc -1 w lp, \
        "plots/2d_stencil/a64fx-new/196608/a64fx_ndouble" using 1:($3/1000) title "vector" lw 2 lt 7 lc -1 w lp, \
        "plots/2d_stencil/a64fx-new/196608/aarch64_bandwidth.dat" using 1:($2/16000) title "Expected peak Max" lw 2 lt 9 lc -1 w lp, \
        "plots/2d_stencil/a64fx-new/196608/aarch64_bandwidth.dat" using 1:($2/24000) title "Expected peak Min" lw 2 lt 9 lc -1 w lp, \

        set size 1,0.15
        set origin 0,-0.05
        unset title
        set key horizontal center right
        unset tics
        unset xlabel
        unset ylabel
        set yrange [0:1]
        plot 2 t 'scalar' lw 2 lc -1 w lp, \
             2 t 'vector' lw 2 lt 7 lc -1 w lp, \
             2 t 'Expected Peak Max' lw 2 lt 9 lc -1 w lp, \
             2 t 'Expected Peak Min' lw 2 lt 9 lc -1 w lp, \

        unset multiplot
    \end{gnuplot}
    \caption{2D stencil: Results for Fujitsu A64FX with a grid size of 8192$\times$196608 iterated over 100 time steps. Expected Peak Max assumes two memory transfers per iteration and Expected Peak Min assumes three memory transfers per iteration.}
    \label{graph:a64fxx}
\end{figure}

We decided to investigate by increasing the grid size. We discovered that 32GB High Bandwidth Memory (HBM) can be
a disadvantage for memory-intensive applications. For instance, our grid
requires 9GB worth of DRAM. A 2D stencil code has two grids, i.e., 18GB worth of
DRAM. We can, therefore, only test grid sizes of up to 1.5x the current size. Our
investigation results suggest that there are no performance benefits (see
Figure~\ref{graph:a64fxx}) in increasing grid size. This means that HPX is able to
take advantage of the underlying parallelism.

\begin{table}[h!]
  \centering
  \caption{Hardware Counters for Fujistu FX1000 A64FX}
  \label{tab:perf-a64fx}
  \begin{tabular}{l|lll}
    \toprule
    \textbf{Data Type}       & \textbf{Instruction} & \textbf{Frontend Stalls} & \textbf{Backend Stalls}  \\ \hline
    \midrule
    Float                   & 1.284$\times 10^{10}$    &  3.801$\times 10^{8}$  & 9.43$\times 10^{9}$ \\ \hline
    Vector Float            & 1.496$\times 10^{10}$    &  2.918$\times 10^{8}$  & 8.003$\times 10^{9}$ \\ \hline
    Double                  & 2.299$\times 10^{10}$    &  3.86$\times 10^{8}$  & 1.871$\times 10^{10}$    \\ \hline
    Vector Double           & 2.956$\times 10^{10}$    &  3.56$\times 10^{8}$  & 1.443$\times 10^{10}$    \\ \hline
    \bottomrule
  \end{tabular}
\end{table}

Another interesting observation is the better performance compared to the
expected peak assuming three memory transfers per iteration. We claimed that the
caches were large enough to accommodate three rows resulting in three memory transfer
per iteration. With A64FX, we observe that the results do not follow the scheme.
Due to large sized cache lines, we see cache benefits resulting in better
Arithmetic Intensity. We witness results equivalent to cache blocking version of
2D stencil up to 32 cores. A cache blocked version of 2D stencil essentially
reduces the number of memory transfers per iteration, in our case, by one. This
results in a 49\% performance boost over the previously expected results.

The significant increase in performance going from half saturated NUMA domain to
a fully saturated NUMA domain can be explained using the same principles as described
for HiSilicon Hi1616. We use a 512-bit SVE vector length for
benchmarking the 2D stencil. From the results, it is clear that no significant improvements are
achieved by explicitly vectorizing the code. The improvements are anywhere from 5\%
to 15\%. Table~\ref{tab:perf-a64fx} provides some insights. Firstly, glancing at the
instruction count, we observe that GCC does a better job of optimizing the instruction
count than our explicitly vectorized code. The cache miss counts were very similar
for auto vectorized and explicitly vectorized codes as well. What differs visibly are the
CPU stall counts. This means that explicit vectorization is helping to relieve stress from
the memory controllers. Due to significant reductions in CPU stalls for vectorized codes,
we get marginally better results.



\begin{figure}[tb]
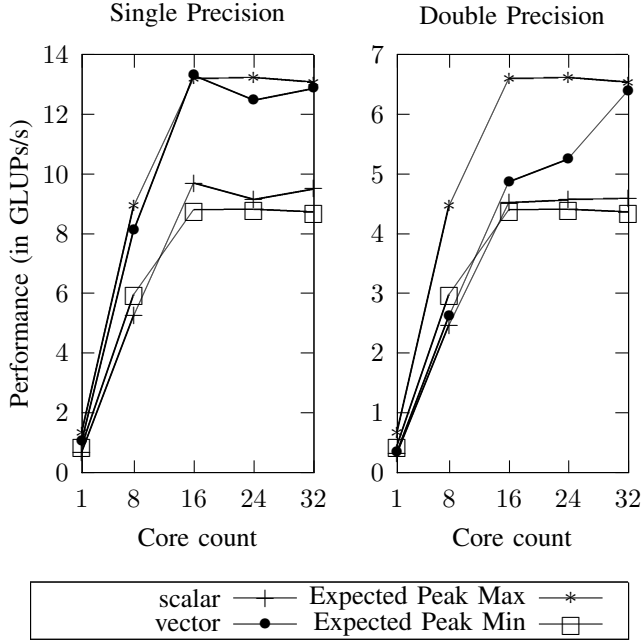

    \centering
    \captionsetup{justification=centering}
    \begin{gnuplot}[terminal=latex]
        set terminal latex rotate size 3.5in,3.8in

        set ylabel "Performance (in GLUPs/s)"
        set xlabel "Core count"

        set size 1,1
        set multiplot layout 2,2 title "2D Stencil: Marvell ThunderX2"

        unset key

        set size 0.53,0.8
        set origin 0,0.1

        set title "Single Precision"

        set xtics (1, 8, 16, 24, 32)

        plot \
        "plots/2d_stencil/x2/x2_float" using 1:($3/1000) title "scalar" lw 2 lc -1 w lp, \
        "plots/2d_stencil/x2/x2_nfloat" using 1:($3/1000) title "vector" lw 2 lt 7 lc -1 w lp, \
        "plots/2d_stencil/x2/aarch64_bandwidth.dat" using 1:($2/8000) title "Expected peak Max" lw 2 lt 9 lc -1 w lp, \
        "plots/2d_stencil/x2/aarch64_bandwidth.dat" using 1:($2/12000) title "Expected peak Min" lw 2 lt 9 lc -1 w lp, \

        set size 0.47,0.8
        set origin 0.53,0.1

        set title "Double Precision"

        unset ylabel

        set xtics (1, 8, 16, 24, 32)

        plot \
        "plots/2d_stencil/x2/x2_double" using 1:($3/1000) title "scalar" lw 2 lc -1 w lp, \
        "plots/2d_stencil/x2/x2_ndouble" using 1:($3/1000) title "vector" lw 2 lt 7 lc -1 w lp, \
        "plots/2d_stencil/x2/aarch64_bandwidth.dat" using 1:($2/16000) title "Expected peak Max" lw 2 lt 9 lc -1 w lp, \
        "plots/2d_stencil/x2/aarch64_bandwidth.dat" using 1:($2/24000) title "Expected peak Min" lw 2 lt 9 lc -1 w lp, \

        set size 1,0.15
        set origin 0,-0.05
        unset title
        set key horizontal center right
        unset tics
        unset xlabel
        unset ylabel
        set yrange [0:1]
        plot 2 t 'scalar' lw 2 lc -1 w lp, \
             2 t 'vector' lw 2 lt 7 lc -1 w lp, \
             2 t 'Expected Peak Max' lw 2 lt 9 lc -1 w lp, \
             2 t 'Expected Peak Min' lw 2 lt 9 lc -1 w lp, \

        unset multiplot
    \end{gnuplot}
    \caption{2D stencil: Results for Marvell ThunderX2 with a grid size of 8192$\times$131072 iterated over 100 time steps. Expected Peak Max assumes two memory transfers per iteration and Expected Peak Min assumes three memory transfers per iteration.}
    \label{graph:x2}
\end{figure}

For Marvell ThunderX2, we see a similar behavior as observed in A64FX. Single
precision performance is taking complete advantage of a large cache line. Cache
blocking behavior can be observed, leading to an additional 49\% performance
boost over an implementation assuming three memory transfers per iteration. With
doubles, we see contradictory behavior. At lower core count, it behaves optimally
according to our assumed arithmetic intensity. At 16 cores and above, the
behavior changes to an arithmetic intensity of 1/8 and 1/16 for floats and
doubles, respectively. This leads to an improvement in performance. Using hardware
stall counters, we found that the number of backend and frontend stalls for explicitly
vectorized code that shows this interesting switch reduced
by about 40\% from $2.353\times 10^{10}$ and $1.144\times 10^{8}$ (for auto vectorization)
to $1.577\times 10^{10}$ and $7.867\times 10^{7}$ (for explicit vectorization),
respectively, at a core count of 32, leading to a significant increase in performance. However, the total number
of instructions and cache misses were similar for both auto vectorized
and explicitly vectorized codes. Unfortunately, we were not able to identify the reason
behind this interesting switch and it remains an open question.

\begin{table}[h!]
  \centering
  \caption{Hardware Counters for Marvell ThunderX2}
  \label{tab:perf-tx2}
  \begin{tabular}{l|lll}
    \toprule
    \textbf{Data Type}       & \textbf{Instruction} & \textbf{L2 Cache Misses} & \textbf{Backend Stalls}  \\ \hline
    \midrule
    Float                   & 4.039$\times 10^{10}$    &  1.811$\times 10^{9}$  & 1.522$\times 10^{10}$   \\ \hline
    Vector Float            & 4.394$\times 10^{10}$    &  1.69$\times 10^{9}$  & 6.437$\times 10^{9}$ \\ \hline
    Double                  & 8.065$\times 10^{10}$    &  5.716$\times 10^{9}$  & 3.298$\times 10^{10}$  \\ \hline
    Vector Double           & 8.756$\times 10^{10}$    &  6.055$\times 10^{9}$  & 2.826$\times 10^{10}$ \\ \hline
    \bottomrule
  \end{tabular}
\end{table}

Effects of adding explicit vectorization are significant. While GCC is able to
auto vectorize the code well (see Table~\ref{tab:perf-tx2}), we notice that the implementation is
rather unoptimized. We observe a decrease in L2 cache misses of about 10\% (No visible difference in L1 cache misses, hence not reported). The
number of backend stalls observed in the case of GCC is considerably
higher compared to an explicitly vectorized code. This means that outstanding load/store
instructions with explicit vectorization is noticeably lower than an auto
vectorized code. In this regard, we believe that ThunderX2 microarchitecture behaves
similar to Cortex-A72 where we saw that cache misses differed by only 15\%, but the
resulting performance gap was upwards of 50\%. We see significant improvements by
utilizing explicit vectorization. These improvements were consistently within 50-60\%
for floats and up to 40\% for doubles. The results also look nearly optimal for the
given memory bandwidth.


\section{Conclusion}

This paper gives a first overview of the performance of AMTs running at scale on
a production HPC system that is based on Arm processors. Our experience of
porting HPX and the benchmarks to Arm processors was mostly straightforward. We
faced a few issues when building for SVE types. Arm Compilers implement SVE
using \lstinline{__sizeless_struct} making it impossible to wrap an SVE type to the
custom container as they have no size at compile time. Currently, only GCC allows the
commandline flag to pass SVE vector length with the \lstinline{-msve-vector-bits}.
However, this comes at the cost of SVE type portability. Further development is required to
integrate custom containers to work with \lstinline{__sizeless_struct} to make
it easier for application developers to port their application to Arm.

We demonstrate that the application scales both on-node and distributed. We
found that performance on Arm processors is as good or better than their x86
brethren. For the 1D stencil, all processors except Kunpeng 916 showed good scaling
results. In the case of Kunpeng 916, the poor interconnect network is to be blamed.
For the 2D stencil, we observed that processors with large cache lines showed
inherent cache blocking benefits (without explicit implementation). This
resulted in about a 50\% performance boost over the expected results. We also
observed that explicit vectorization can improve the performance
significantly for ThunderX2 and Kunpeng 916 due to considerably lower CPU stall
counts when compared with auto vectorized codes. For A64FX, we did not observe any
visible performance benefits by employing explicit vectorization.

\section{Acknowledment}

We  would  like  to  thank  Huawei  for  making  the  JUAWEI cluster  hardware
available  at  JSC  as  well  as  Fujitsu  for allowing us to participate in the
A64FX early access program. Funding for parts of this work is received from the
European Commission H2020 program under Grant Agreement 779877(Mont-Blanc
2020).

We are grateful for the insights provided by Dr. Thomas Heller (Exasol) and
Prof. Hartmut Kaiser (LSU, USA) in analyzing and optimizing our benchmarks.

\bibliographystyle{IEEEtran}
\bibliography{reference}

\end{document}